\documentclass[10pt,superscriptaddress,nofootinbib,longbibliography]{revtex4-2}

\usepackage[Symbol]{upgreek}
\usepackage{amsmath}
\usepackage{amssymb}
\usepackage{enumitem}
\usepackage{xcolor} 
\usepackage{tensor}

\usepackage{hyperref}
\usepackage{booktabs}
\hypersetup{
    colorlinks=true,
    linkcolor=blue,
    filecolor=magenta,      
    urlcolor=cyan,
}

\newcommand{\red}[1]{\textcolor{red}{#1}}

\begin{document}

\title{The Gravitational Aspect of Information: \\ \large The Physical Reality of Asymmetric ``Distance"}

\author{Tomoi Koide}
\email{tomoikoide@gmail.com,koide@if.ufrj.br}
\affiliation{Instituto de F\'{\i}sica, Universidade Federal do Rio de Janeiro, 
21941-972, Rio de Janeiro, RJ, Brazil}
\author{Armin van de Venn}
\email{Armin.vandevenn@gmx.net}
\affiliation{Frankfurt Institute for Advanced Studies (FIAS), Frankfurt am Main, Germany}
\affiliation{Physics Department, Goethe University, Max-von-Laue-Str.~1, 60438 Frankfurt am Main, Germany}

\begin{abstract}
We show that when a Brownian bridge is physically constrained to satisfy a canonical condition, its time evolution exactly coincides with an m-geodesic on the statistical manifold of Gaussian distributions. This identification provides a direct physical realization of a geometric concept in information geometry. It implies that purely random processes evolve along informationally straight trajectories, analogous to geodesics in general relativity. Our findings suggest that the asymmetry of informational ``distance" (divergence) plays a fundamental physical role, offering a concrete step toward an equivalence principle for information.
\end{abstract}

\maketitle

\section{Introduction}

In quantum information and quantum thermodynamics, it is often necessary to quantify the similarity of two quantum states. If one interprets this notion of similarity as a literal distance, it is intuitively expected to be symmetric under the exchange of its arguments. Indeed, foundational measures such as fidelity,\footnote{The fidelity between two quantum states $\rho$ and $\sigma$ is defined as $F(\rho, \sigma) = \left( \text{Tr} \sqrt{\sqrt{\rho}\sigma\sqrt{\rho}} \right)^2$.} the Bures distance,\footnote{The Bures distance is derived from the fidelity and is given by $D_B(\rho, \sigma) = \sqrt{2(1 - \sqrt{F(\rho, \sigma)})}$.} and the trace distance\footnote{The trace distance is defined as $D_{tr}(\rho, \sigma) = \frac{1}{2} \text{Tr}|\rho - \sigma|$, where $|A| = \sqrt{A^\dagger A}$.} all satisfy this symmetric property \cite{nielsen,Bengtsson}. These measures are foundational as they provide physically meaningful ways to quantify the similarity between quantum states.

However, an alternative concept, born from information theory, also exists. This is the notion of divergence, a quantity that serves as an information-theoretic analogue to distance by measuring the gap between two statistical distributions. A crucial difference, however, is that most divergences are inherently asymmetric. This asymmetry often arises because the divergence is defined via an expectation value, which requires choosing one of the two distributions as the reference for the calculation. This apparent violation of a fundamental property of distance may lead one to question their utility in physical theories. The central argument of this paper is that this very asymmetry, rather than being a flaw, is a crucial feature that reveals a deeper connection to physical processes.

To demonstrate this argument convincingly, we will turn to the well-established framework of classical information geometry \cite{amari1985,amari2000,ay2017}. 
This field, where the geometric structure is uniquely determined by the Fisher information metric and the cubic tensor, provides an ideal and unambiguous laboratory to explore the physical meaning of an asymmetric structure. 
By proving our argument in this clear setting, we aim to build a strong case for the potential benefits of applying similar perspectives to the more complex quantum realm, where even the choice of a fundamental metric is non-trivial.

The idea of applying differential geometry to statistics has a rich history. It can be traced back to the work of Hotelling and was later formalized by Rao, who introduced the Fisher information as a metric tensor~\cite{hotelling1930,rao1945}. Chentsov subsequently proved the uniqueness of this metric~\cite{chentsov1982}, and 
Efron related its curvature to statistical efficiency~\cite{efron1975}, while Dawid was the first to explicitly connect this curvature to the deeper structure of dual affine connections~\cite{dawid1975}.
It was, however, the work of Amari that systematically established the field of information geometry, crucially introducing the dual structure of $(\alpha)$-connections that effectively handles the asymmetry central to our discussion~\cite{amari1985,amari2000,ay2017}.

Within this robust framework, 
we will show how an asymmetric divergence can be understood as a natural generalization of the symmetric squared Euclidean distance. 
This result itself is already well-known in the field of information geometry \cite{amari2000}, 
but it provides a necessary foundation for our arguments.
Building upon this framework, we next turn to the notion of geodesics. We identify a physical process that exactly realizes an m-geodesic, thereby giving a concrete physical meaning to what has so far been a purely geometric concept. Specifically, we demonstrate that the time evolution of a Brownian bridge, when constrained by a \textit{canonical} condition, coincides exactly with an m-geodesic on the statistical manifold of Gaussian distributions.

This correspondence allows for a powerful new interpretation: just as a geodesic in general relativity describes the motion of a free particle subject only to gravity, the evolution of a purely random process can be seen as following a geodesic trajectory on a statistical manifold. 
This correspondence provides a novel physical interpretation of the nature of stochastic processes, revealing a fundamental geometric principle that governs their behavior.

Throughout this paper, we will use the Einstein summation convention, where summation is implied for repeated indices.

The remainder of this paper is organized as follows. Section \ref{sec:ig} briefly reviews the general framework of information geometry. 
The specific application to the exponential family is discussed in Sec. \ref{sec:ef}. 
Section \ref{sec:diver} presents that the Bregman divergence is a generalization of the Euclidean distance, a fact that, while known to experts, is foundational to our argument. 
Our main result is shown in Sec.\ \ref{sec:bb}, establishing the correspondence between m-geodesics and the canonical Brownian bridge. By directly applying the dual coordinate framework developed in the earlier sections, we demonstrate that the time evolution of this stochastic process coincides exactly with an m-geodesic. This identification provides, for the first time, a concrete physical meaning to what has so far been a purely geometric concept.
Section \ref{sec:conc} is devoted to concluding remarks.

\section{General Framework of Information Geometry}
\label{sec:ig}

Before focusing on a specific example, we briefly introduce the general concepts of information geometry. The central idea is to treat a family of probability distributions as a geometric space, a ``statistical manifold," and to analyze its properties using the tools of differential geometry.

\subsection{Statistical Manifold}

A family of probability distributions $p(x;\theta)$ parametrized by $\theta = (\theta^1, \dots, \theta^n)$ forms a statistical manifold, where each point $\theta$ corresponds to a single distribution. To measure the ``distance" between two infinitesimally close distributions, we introduce a metric tensor. The natural choice, unique up to a constant factor, is the Fisher information matrix $g_{ij}(\theta)$ \cite{chentsov1982}. 
The components of the Fisher information matrix are defined as the covariance of the score vector (the gradient of the log-likelihood):
\begin{equation}
    g_{ij}(\theta) = E_{\theta}\left[ \left(\frac{\partial}{\partial\theta^i} \log p(x;\theta)\right) \left(\frac{\partial}{\partial\theta^j} \log p(x;\theta)\right) \right] \, ,
\end{equation}
where $E_{\theta}[\cdot]$ is the expectation value with $p(x;\theta)$.
A statistical manifold equipped with this Fisher metric becomes a Riemannian manifold.

The geometry of a statistical manifold is richer than that of a standard Riemannian manifold. This additional structure is captured by a symmetric tensor of order 3, the cubic tensor $T_{ijk}(\theta)$, defined by the third-order correlations of the score vector:
\begin{equation}
    T_{ijk}(\theta) = E_{\theta}\left[ \left(\frac{\partial \log p(x;\theta)}{\partial\theta_i}\right) \left(\frac{\partial \log p(x;\theta)}{\partial\theta_j}\right) \left(\frac{\partial \log p(x;\theta)}{\partial\theta_k}\right) \right] \, .
\end{equation}
The presence of this cubic tensor allows for the definition of a continuous family of affine connections, known as $(\alpha)$-connections. 
These connections are assumed to be torsion-free, meaning their connection coefficients are symmetric in the lower indices. They are defined by modifying the Levi-Civita connection:
\begin{equation}
    {}^{(\alpha)}{\Gamma^{i}}_{jk}(\theta) = \tensor{\mathring{\Gamma}}{^{i}_{jk}}(\theta) - \frac{\alpha}{2}g^{il}T_{jkl}(\theta) \, , \label{eqn:alpha-conn}
\end{equation}
where $\tensor{\mathring{\Gamma}}{^{i}_{jk}}$ are the Christoffel symbols of the Levi-Civita connection and $\alpha$ is any real constant.

It is important to note that the $(\alpha)$-connection is a genuine affine connection. 
As such, its coefficients transform according to the established rule for connections under a change of coordinates. 
Let us consider a coordinate transformation from $\theta = (\theta^i)$ to $\xi = (\xi^a)$. 
Using the definition (\ref{eqn:alpha-conn}), the connection coefficients in the new coordinates, ${}^{(\alpha)}\tilde{\Gamma^{c}}_{ab}$, are related to the original coefficients by the well-known relation for connections,
\begin{equation}
   {}^{(\alpha)} {\tilde{\Gamma}}\indices{^{c}_{ab}}(\xi) = \frac{\partial\xi^c}{\partial\theta^k} \frac{\partial\theta^i}{\partial\xi^a} \frac{\partial\theta^j}{\partial\xi^b} {}^{(\alpha)}{\Gamma^k}_{ij}(\theta) + \frac{\partial\xi^c}{\partial\theta^k} \frac{\partial^2\theta^k}{\partial\xi^a \partial\xi^b} \, .
\end{equation}

In this geometric framework, tangent vectors are represented as directional derivative operators. In the local coordinate system $\{\theta \}$, the set of partial derivative operators $\{\partial_i := \partial/\partial\theta^i\}$ forms a basis for the tangent space at each point. Any tangent vector field $X$ can therefore be expressed as a linear combination of these basis vectors: $X = X^i \partial_i$, where $X^i$ are the component functions.

Each $(\alpha)$-connection defines a notion of parallel transport and differentiation on the manifold. The covariant derivative of a vector field $V$ along another vector field $Z$, denoted as ${}^{(\alpha)}\nabla_Z V$, is given in local coordinates by
\begin{equation}
    ({}^{(\alpha)}\nabla_k V)^i := \partial_k V^i + {}^{(\alpha)}\tensor{\Gamma}{^i_{lk}} V^l \, .
\end{equation}
A crucial feature of this structure is the concept of duality. The $(-\alpha)$-connection is said to be dual to the $(\alpha)$-connection with respect to the Fisher metric. This duality is expressed by a generalized product rule that relates the directional derivative of an inner product to the two connections. For any vector fields $X, Y, Z$:
\begin{equation}
    Z\langle X,Y\rangle 
    = \langle {}^{(\alpha)}\nabla_Z X,Y\rangle + \langle X, {}^{(-\alpha)}\nabla_Z Y\rangle \, ,
\end{equation}
where $\langle X,Y\rangle := g_{ij} X^i Y^j$ is the inner product induced by the Fisher metric, and $Z\langle X,Y\rangle$ is the directional derivative of this scalar function along $Z$. This relation can be seen as a generalization of the product rule for derivatives in Euclidean space. It dictates that the change in the inner product (which determines lengths and angles) is split between the two dual connections. In component form, this duality implies the following relation for the metric tensor:
\begin{equation}
    {}^{(\alpha)}\nabla_{k}g_{ij} = \left({}^{(-\alpha)}\Gamma\indices{_{ijk}} - {}^{(\alpha)}\Gamma\indices{_{ijk}}\right) \, ,
\end{equation}
where ${}^{(\alpha)}\Gamma_{ijk} := g_{il} {}^{(\alpha)}\tensor{\Gamma}{^l_{jk}}$ is the fully covariant form of the connection coefficients. Physically, this relation signifies that the $(\alpha)$-connection is not metric-compatible, meaning the length of a vector is generally not preserved under parallel transport using only an $(\alpha)$-connection.

Finally, each connection defines its own set of ``straight lines," or geodesics. A trajectory $\theta(t)$ is called an $(\alpha)$-geodesic if it satisfies the geodesic equation:
\begin{equation}
    \frac{\mathrm{d}^2\theta^i(t)}{\mathrm{d}t^2} + {}^{(\alpha)}\tensor{\Gamma}{^i_{jk}} \frac{\mathrm{d}\theta^j(t)}{\mathrm{d}t} \frac{\mathrm{d}\theta^k(t)}{\mathrm{d}t} = 0 \, ,
\end{equation}
where $t$ is an affine parameter. These dual connections and their corresponding geodesics are fundamental to understanding the geometry of statistical models.

\subsection{Divergence}

A divergence $D(\theta_p\|\theta_q)$ is a function that quantifies the difference between two distributions with $\theta_p$ and $\theta_q$. While not a true distance, it must satisfy several key properties analogous to distance:
\begin{enumerate}
    \item $D(\theta_p\|\theta_q) \ge 0$ for all points $\theta_p, \theta_q$ on the manifold. 
    \item $D(\theta_p\|\theta_q) = 0$ if and only if $\theta_p=\theta_q$. 
    \item For two infinitesimally close distributions with parameters $\theta$ and $\theta+\mathrm{d}\theta$, the divergence is related to the Fisher metric as:
    \begin{equation}
        D(\theta+\mathrm{d}\theta\|\theta) = \frac{1}{2}g_{ij}(\theta)\mathrm{d}\theta^i \mathrm{d}\theta^j + O(\|\mathrm{d}\theta\|^3) \, .
    \end{equation}
    The factor of $1/2$ in the leading term is a convention chosen so that the divergence locally matches the squared Euclidean distance, $ds^2 = g_{ij} \mathrm{d}\theta^i \mathrm{d}\theta^j$ for the Gaussian statistical manifold with $^{(\alpha)}\nabla = ^{(-\alpha)}\nabla$.
\end{enumerate}
From a given divergence function that meets these criteria, one can recover the metric and the dual connections through differentiation. 
We are particularly interested in the behavior of statistical manifolds to which exponential functions belong.
Then, using the Bregman divergence which we introduce later, the Fisher metric is obtained by:
\begin{equation}
    g_{ij}(\theta) = - \left[ \frac{\partial}{\partial(\theta_p)^i} \frac{\partial}{\partial(\theta_q)^j} D_{BR} (\theta_p\|\theta_q) \right]_{\theta_p=\theta_q=\theta}\, .
\end{equation}
Similarly, the connection components (in covariant form) of the two fundamental dual connections are derived from third-order derivatives of the divergence. The first is the e-connection (exponential connection, for $\alpha=1$):
\begin{equation}
    ^{(e)}\Gamma_{kij}(\theta) := 
    g_{kl}(\theta) \,\, {}^{(e)}{\Gamma^l}_{ij}(\theta)
    =
    - \left[ \frac{\partial}{\partial(\theta_p)^i} \frac{\partial}{\partial(\theta_p)^j} \frac{\partial}{\partial(\theta_q)^k} D_{BR} (\theta_p\|\theta_q) \right]_{\theta_p=\theta_q=\theta} \label{eqn:e-conn} \, .
\end{equation}
The second is its dual partner, the m-connection (mixture connection, for $\alpha=-1$):
\begin{equation}
    ^{(m)}\Gamma_{kij}(\theta) := 
    g_{kl}(\theta)\,\, {}^{(m)}{\Gamma^l}_{ij}(\theta)
    =
    - \left[ \frac{\partial}{\partial(\theta_q)^i} \frac{\partial}{\partial(\theta_q)^j} \frac{\partial}{\partial(\theta_p)^k} D_{BR}(\theta_p\|\theta_q) \right]_{\theta_p=\theta_q=\theta} \, .
\end{equation}
The definition for more general $(\alpha)$-connection is shown in Ref.\ \cite{amari2000}.

\section{A Key Example: The Exponential Family}
\label{sec:ef}

We now focus on the exponential family of probability distributions. This family is not only a setting where the general structure of information geometry becomes remarkably clear, but it is also the most important statistical manifold from a physical standpoint. 
This is because it is deeply connected to the thermal equilibrium distributions, which lie at the heart of statistical mechanics \cite{jaynes1957}.

An exponential family is a set of probability distributions whose probability density function can be written in the form:
\begin{equation}
    p(x;\theta) = \exp\left((\theta)^i F_{i}(x) - \psi(\theta)\right) \, ,
\end{equation}
where $\theta$ are the natural parameters and $\psi(\theta)$ is a function called the potential. For this family, the Fisher metric $g_{ij}$ and the cubic tensor $T_{ijk}$ are given by the second and third derivatives of this potential:
\begin{align}
    g_{ij}(\theta) &= \frac{\partial^2 \psi(\theta)}{\partial\theta^i \partial\theta^j} \label{eq:g_from_psi} \\
    T_{ijk}(\theta) &= \frac{\partial^3 \psi(\theta)}{\partial\theta^i \partial\theta^j \partial\theta^k} \, .
\end{align}
Equation \eqref{eq:g_from_psi} proves that $\psi(\theta)$ is a convex function, 
as its Hessian matrix is the Fisher metric which is positive semi-definite by definition.

The convexity of $\psi(\theta)$ allows us to define a Legendre transformation, which introduces a dual coordinate system $\eta$, the expectation parameters:
\begin{equation}
    \eta_{i} = \frac{\partial\psi(\theta)}{\partial\theta^{i}} \, .
\end{equation}
This transformation also defines a dual potential $\phi(\eta)$ via the Legendre transformation:
\begin{equation}
    \phi(\eta) = \max_{\theta} \{ (\theta)^i \eta_i - \psi(\theta) \} \, .
\end{equation}
The pair of dual coordinates $(\theta, \eta)$ is a central feature of the exponential family.

Furthermore, the potential $\psi(\theta)$ naturally defines the Bregman divergence:
\begin{equation}
    D_{BR}(\theta_p\|\theta_q) := \psi(\theta_p) - \psi(\theta_q) - (\theta_p - \theta_q)^i \frac{\partial\psi(\theta_q)}{\partial(\theta_q)^i} \, .
\end{equation}
Using this divergence, we can explicitly calculate the e-connection coefficients from the general formula in Eq.\ (\ref{eqn:e-conn}). Differentiating $D_{BR}$ with respect to its arguments yields:
\begin{equation}
    ^{(e)}\Gamma_{kij}(\theta) =-\left[ \frac{\partial}{\partial(\theta_p)^i} \frac{\partial}{\partial(\theta_p)^j} \frac{\partial}{\partial(\theta_q)^k} D_{BR}(\theta_p\|\theta_q) \right]_{\theta_p=\theta_q=\theta} = 0 \, .
\end{equation}
Since the e-connection coefficients vanish in the $\theta$-coordinates, the corresponding geodesics (e-geodesics) are straight lines in this coordinate system.

Dually, we can define the dual Bregman divergence, $D_{BR}^*$, based on the dual potential $\phi(\eta)$:
\begin{equation}
    D_{BR}^*(\eta_p\|\eta_q) := \phi(\eta_p) - \phi(\eta_q) - (\eta_p - \eta_q)_i \frac{\partial\phi(\eta_q)}{\partial(\eta_q)_i} 
    = D_{BR} (\theta_q \| \theta_p)
    \, .
\end{equation}
From this, one can see that the role of $p$ and $q$ is inverted in moving from the natural coordinate to the dual coordinates. Thus the definition of the e- and m-connections is inverted in the dual coordinates.
The m-connection coefficients are then defined analogously using this dual divergence:
\begin{equation}
    ^{(m)}\tilde{\Gamma}^{kij}(\eta) := - \left[ \frac{\partial}{\partial(\eta_p)_i} \frac{\partial}{\partial(\eta_p)_j} \frac{\partial}{\partial(\eta_q)_k} D_{BR}^*(\eta_p\|\eta_q) \right]_{\eta_p=\eta_q=\eta} = 0 \, .
    \label{eqn:m-conn-coef}
\end{equation}
A calculation parallel to that for the e-connection confirms that these coefficients also vanish. 
This proves that m-geodesics are straight lines in the $\eta$-coordinate system. 
A manifold exhibiting this property is said to possess a dually flat structure.
Because the Gaussian manifold is dually flat, both e- and m-geodesics are globally straight in their respective coordinates, ensuring that this correspondence holds globally rather than locally.

\section{Interrelations of Divergences and the Euclidean Limit}
\label{sec:diver}

In this section, we show how the various divergences are related and, crucially, how they connect back to the familiar Euclidean distance \cite{amari2000}.

\subsection{Relations between Divergences}

First, we introduce the canonical divergence, which is defined using both dual potentials:
\begin{equation}
    D_{CA}(\theta_p\|\eta_q) := \psi(\theta_p) + \phi(\eta_q) - (\theta_p)^i (\eta_q)_i\, .
\end{equation}
This connects the geometric structure to the more familiar information-theoretic measures. By substituting the definitions of the Legendre transform and the Bregman divergence, one can prove the following important set of identities for the exponential family:
\begin{align}
    D_{CA}(\theta_p\|\eta_q) = D_{BR}(\theta_p\|\theta_q) 
    = D_{BR}^*(\eta_q\|\eta_p) 
    = D_{KL}(p(\theta_q)\|p(\theta_p)) \, .
\end{align}
Here, $D_{KL}(p(\theta_p)\|p(\theta_q))$ is the well-known Kullback-Leibler (KL) divergence, defined as the expectation under $p(x;\theta_p)$ of the logarithmic difference between the two probability densities:
\begin{equation}
    D_{KL}(p(\theta_p)\|p(\theta_q)) = E_{\theta_p}\left[ \log \frac{p(x;\theta_p)}{p(x;\theta_q)} \right] \, .
\end{equation}
These identities reveal that the Bregman and canonical divergences derived from the potentials are not new quantities, but are in fact equivalent to the fundamental KL divergence. 

\subsection{The Euclidean Limit: The Self-Dual Case}

Now we can demonstrate how this generalized geometric structure contains the familiar Euclidean distance as a special case \cite{amari2000}. This occurs when the manifold is self-dual, which corresponds to the condition that the cubic tensor vanishes, i.e., $T_{ijk}=0$.

The vanishing of the cubic tensor implies that the potential $\psi(\theta)$ must be a polynomial of at most degree 2 in $\theta$. 
Consequently, the metric components, $g_{ij}(\theta) = \partial_i \partial_j \psi(\theta)$, must be constant. 
Ignoring irrelevant constant and linear terms, the potential can be written as a simple quadratic form:
\begin{equation}
    \psi(\theta) = \frac{1}{2}g_{kl}\theta^k \theta^l \, .
\end{equation}
Under this condition, the dual coordinates $\eta_i$ become linearly related to the $\theta$ coordinates:
\begin{equation}
    \eta_i = \frac{\partial \psi}{\partial \theta^i} = g_{ik} \theta^k \, .
\end{equation}
Let us now evaluate the Bregman divergence for this self-dual case. Substituting the quadratic potential into the definition of $D_{BR}(\theta_q\|\theta_p)$ gives:
\begin{align}
    D_{BR}(\theta_q\|\theta_p) 
    &= \frac{1}{2}g_{ij}(\theta_q)^i (\theta_q)^j - \frac{1}{2}g_{ij}(\theta_p)^i (\theta_p)^j - ((\theta_q)^i - (\theta_p)^i)g_{ik}(\theta_p)^k \nonumber \\
    &= \frac{1}{2}g_{ij}\left( (\theta_q)^i - (\theta_p)^i \right) \left( (\theta_q)^j - (\theta_p)^j \right) \, .
\end{align}
This explicit recovery of Euclidean geometry is a foundational result in information geometry, providing a strong justification that divergence can be considered a natural generalization of distance \cite{amari2000}.

\section{Geodesics and the Brownian Bridge}
\label{sec:bb}

The true power of this geometric perspective becomes apparent when we connect it to physical processes. We will now show that the geodesics on the statistical manifold of Gaussian distributions are directly related to the time evolution of a physical system.

\subsection{The Gaussian Statistical Manifold}
Let us consider the statistical manifold of one-dimensional Gaussian distributions as a concrete example. A Gaussian distribution is specified by its mean $\mu$ and variance $\sigma^2$:
\begin{equation}
    p(x;\mu,\sigma^2) = \frac{1}{\sqrt{2\pi\sigma^2}} \exp\left(-\frac{(x-\mu)^2}{2\sigma^2}\right) \, .
\end{equation}
This can be written in the exponential family form $p(x;\theta) = \exp((\theta)^i F_i(x) - \psi(\theta))$ by identifying $F(x)=(x, x^2)$ and the natural coordinates $(\theta^1, \theta^2)$ as:
\begin{equation}
    (\theta^1, \theta^2) = \left(\frac{\mu}{\sigma^2}, -\frac{1}{2\sigma^2}\right)\, .
\end{equation}
From this, we can derive the potential $\psi(\theta)$ in terms of the natural coordinates:
\begin{equation}
    \psi(\theta) = \frac{\mu^2}{2\sigma^2} + \frac{1}{2}\ln(2\pi\sigma^2) 
    = -\frac{(\theta^1)^2}{4\theta^2} + \frac{1}{2}\ln(\pi) - \frac{1}{2}\ln(-\theta^2) \, .
\end{equation}
The dual coordinates, $\eta_i$, which are the gradients of the potential, correspond to the first and second moments of the distribution:
\begin{align}
    \eta_1 &= \frac{\partial \psi}{\partial \theta^1} = -\frac{\theta^1}{2\theta^2} = \mu \, ,\\
    \eta_2 &= \frac{\partial \psi}{\partial \theta^2} = \frac{(\theta^1)^2}{4(\theta^2)^2} - \frac{1}{2\theta^2} = \mu^2 + \sigma^2 \, .
\end{align}

Although the (Bregman) divergence serves as the fundamental measure of asymmetric informational distance as discussed in Sec. IV, its explicit algebraic evaluation is not strictly necessary for determining the information-geometric quantities of the system. As established in Sec. \ref{sec:ef}, the entire geometric structure of the Gaussian family governing the trajectory, including the metric and connection coefficients, is completely encapsulated by the potential $\psi(\theta)$ and its dual coordinates $\eta_i$. Therefore, rather than calculating the divergence itself, we can proceed directly to deriving the geodesic equations using these exact coordinates.

\subsection{The Brownian Bridge as an m-Geodesic}


\subsubsection{Derivation of the Physical Trajectory}

A Brownian bridge (or pinned Brownian motion) describes a diffusing particle that starts at position $x_a$ at time $t=t_a$ and is conditioned to end at position $x_b$ at time $t=t_b$. The process can be described by the following stochastic differential equation (SDE) \cite{ezawa}:
\begin{equation}
    \mathrm{d}X(t) = \frac{x_b - X(t)}{t_b-t}\mathrm{d}t + \sqrt{2D}\,\mathrm{d}B_t \, ,
    \label{eqn:sde_BB}
\end{equation}
where $D$ is the diffusion coefficient and $\mathrm{d}B_t$ is a Wiener process. To solve this, one can introduce an auxiliary variable $Y(t) = (X(t) - x_b)/(t_b-t)$. The SDE for $Y(t)$ simplifies to
\begin{equation}
    \mathrm{d}Y(t) = \frac{\sqrt{2D}}{t_b-t}\mathrm{d}B_t \, .
\end{equation}
Since this is a linear SDE characterized by the Wiener process, its solution $Y(t)$ follows a normal distribution. 
Consequently, the probability distribution for the original process, $P(X(t)=x)$, is also a normal distribution:
\begin{equation}
    P(X(t) =x) = n(x- \mu_{BB}(t) , \sigma^2_{BB} (t)) \, ,
\end{equation}
where 
\begin{equation}
    n(x,v) = \frac{1}{\sqrt{2\pi v}} e^{-x^2/(2v)} \, . 
\end{equation}
By solving the equations, one finds the mean and variance for $t_a < t < t_b$:
\begin{align}
    \mu_{BB}(t) &=x_a + \frac{t-t_a}{t_b-t_a}(x_b - x_a) \, ,\\
   \sigma^2_{BB} (t) &= 2D \frac{(t-t_a) (t_b - t)}{t_b - t_a} \, .
\end{align}
In this standard formulation of the Brownian bridge, the parameters $t_a, t_b, x_a, x_b,$ and $D$ are all independent.

\subsubsection{Derivation of the Geometric Trajectory}

Next, let us determine the purely geometric trajectory connecting the same two endpoints. 
As rigorously demonstrated for the dually flat structure of the exponential family in Sec. \ref{sec:ef} (specifically, see the discussion surrounding Eq. (\ref{eqn:m-conn-coef})), the vanishing of the m-connection coefficients guarantees that an m-geodesic forms a straight line in the dual $\eta$-coordinate system. 
Applying this fundamental property to our Gaussian manifold, we set the boundary conditions such that the process starts at $x_a$ with zero variance and ends at $x_b$ with zero variance.
The boundary conditions are that the process starts at $x_a$ with zero variance and ends at $x_b$ with zero variance. In the $\eta$-coordinates, 
$(\eta_1(t_a),\eta_2 (t_a)) = (x_a,x^2_a)$ and 
$(\eta_1(t_b),\eta_2 (t_b)) = (x_b,x^2_b)$.
Since the trajectory is linear, $\eta_i(t) = c_i t + d_i$, we can solve for the four coefficients: 
\begin{align}
    c_1 = \frac{x_b-x_a}{t_b-t_a}, \quad &d_1 = \frac{t_b x_a - t_a x_b}{t_b-t_a} \, , \\
    c_2 = \frac{x_b^2-x_a^2}{t_b-t_a}, \quad &d_2 = \frac{t_b x_a^2 - t_a x_b^2}{t_b-t_a} \, .
\end{align}
Substituting these coefficients into the expressions for the mean and variance, $\mu_{\text{geo}}(t) = \eta_1(t)$ and $\sigma^2_{\text{geo}}(t) = \eta_2(t) - (\eta_1(t))^2$, gives the trajectory of the m-geodesic:
\begin{align}
    \mu_{\text{geo}}(t) &= x_a + \frac{t-t_a}{t_b-t_a}(x_b - x_a) \, ,\\
    \sigma^2_{\text{geo}}(t) &= \frac{(x_b-x_a)^2}{(t_b-t_a)^2}(t-t_a)(t_b-t) \, .
\end{align}

We compare the physical and geometric trajectories. 
The evolutions of the mean are identical, $\mu_{BB}(t) = \mu_{\text{geo}}(t)$. However, the variances are equivalent, $\sigma^2_{BB}(t) = \sigma^2_{\text{geo}}(t)$, if and only if the parameters satisfy the condition:
\begin{equation}
   D = \frac{1}{2} \frac{(x_b - x_a)^2}{t_b - t_a} \label{eqn:cond_CBB_new} \, .
\end{equation}
This arises as a mathematical necessity to make the physical and geometric trajectories identical.  
This relation implies that the total displacement $|x_b-x_a|$ is naturally scaled 
by the characteristic displacement of the Brownian motion, $\sqrt{2D(t_b - t_a)}$.
Indeed, if the condition (\ref{eqn:cond_CBB_new}) is satisfied, 
the diffusion coefficient $D$ is always given by the second moment of the displacement 
for any $t_a < t < t_b$,
\begin{align}
   \frac{E_w [(X(t) - x_a)^2]}{t-t_a} 
   = 2D \, ,
\end{align}
where $E_w [\cdot]$ denotes the expectation value for the Wiener process.
This ensures that the mean squared displacement grows linearly with time, making the process statistically self-similar and hence ‘canonical’.
We therefore call a Brownian bridge that satisfies Eq. \eqref{eqn:cond_CBB_new} a canonical Brownian bridge. 
It is also worth mentioning that the infinite limit of $t_b$ of the Brownian bridge is known to reduce to Brownian motion.

With this final piece, we can state our main result:
The time evolution of a canonical Brownian bridge is identical to an m-geodesic trajectory on the statistical manifold.

\section{Concluding remarks}
\label{sec:conc}

In this paper, first, we confirmed that the Bregman divergence (and hence the KL divergence) serves as a generalization of the squared Euclidean distance for an exponential family. 
This result itself is already well-known in the field of information geometry \cite{amari2000}, 
but it provides an important foundation for our arguments by showing that the asymmetric divergence naturally contains the familiar symmetric distance as a special case.

Our second, more significant finding is the deep connection between geodesics and a physical stochastic process. To make this connection precise, we introduced the concept of the ``canonical Brownian bridge": 
a Brownian bridge that satisfies a specific physical condition ensuring its mean squared displacement from the start point scales linearly with time, just as in standard diffusion. 
We then demonstrated that the time evolution of this process corresponds exactly to an m-geodesic on the statistical manifold of Gaussian distributions. 
The m-geodesic represents the trajectory that keeps the informational distance (KL divergence) from the origin minimal at each step, in the sense of a projection \cite{amari2000}. 
Our result therefore implies that the natural random process of the canonical Brownian bridge behaves as if it follows an optimization principle, evolving in a way that is informationally ``straightest".

\begin{table}[t!]
\centering
\caption{Analogy between General Relativity and Information Geometry.}
\label{tab:analogy}
\begin{tabular}{ll}
\toprule
\textbf{General Relativity} & \textbf{Information Geometry} \\
\midrule
Spacetime & Statistical Manifold \\
Free Particle Motion & Purely Random Process \\
& (e.g., Canonical Brownian Bridge) \\
Force  & Information Constraint \\
Equivalence Principle & Equivalence Principle for Information (?) \\
\bottomrule
\end{tabular}
\end{table}

This correspondence allows for a powerful new interpretation reminiscent of Einstein's equivalence principle. 
In general relativity, a free particle (subject only to gravity) follows a geodesic in curved spacetime, and one can always find a local free-falling frame where this motion appears as a straight line. 
Inspired by this, 
we suggest an information-theoretic analogue of the equivalence principle: 
on any given statistical manifold, a process that is "perfectly random", subject to no external forces or biases, will follow a geodesic trajectory. 
This principle reframes randomness not as noise, but as a form of free motion guided by the underlying geometry of information. 
Our work provides the first rigorous proof of this principle in a non-trivial physical system. 
We have shown that for the Gaussian manifold, the ``perfectly random" process is the canonical Brownian bridge, and its trajectory is identical to an m-geodesic. Because this manifold is dually flat, this ''free motion" is not just a local property but a global one. This perspective suggests a geometric foundation for randomness, where deviations from geodesic motion on a general, curved statistical manifold could be described by its intrinsic curvature, quantifying the ``informational forces" that distort the process from pure randomness.
The analogy between general relativity and information geometry is summarized in Table \ref{tab:analogy}.

While the proposed equivalence principle for information is currently a heuristic analogy, our results suggest that a rigorous formulation may be possible in more general statistical manifolds.

Our work finds its place among a growing body of research connecting information geometry and physical dynamics. For instance, Crooks proposed that the concept of thermodynamic length connects the Fisher metric to thermodynamic costs~\cite{crooks2007}, an idea later extended by Ito~\cite{ito2018}. In parallel, other research has linked geodesic trajectories to physical dynamics: Fujiwara and Amari showed that the gradient flow of KL divergence follows a geodesic~\cite{fujiwara1995}, 
and the Schr\"{o}dinger bridge problem can be viewed as finding a geodesic in the space of probability measures~\cite{zambrini_book,leonard2014,Conforti2019}.
Ohara has shown a similar result to us, though there is a crucial distinction: in his work on the porous medium equation, it is the \textit{m-projection} of the solution onto the manifold that traces the geodesic, as the solution itself may lie outside the manifold~\cite{ohara2009}. In our case, the correspondence is more direct: the time evolution of the physical process \textit{itself} constitutes the geodesic trajectory. Taken together, these findings suggest that the identification of a physical process with a geodesic is not an isolated coincidence but a more universal principle.

This successful identification of a physical meaning for geodesics in the classical setting provides a strong motivation to pursue a similar program in the quantum realm. The extension to quantum information geometry is, however, highly non-trivial. 
Unlike the classical case where the Fisher information metric is essentially unique, the non-commutativity of quantum operators gives rise to a multitude of candidates for quantum Fisher metrics. 
A whole family of such metrics, known as monotone metrics, has been studied, including the Symmetric Logarithmic Derivative (SLD) Fisher metric and the Bogoliubov-Kubo-Mori (BKM) metric as notable examples~\cite{petz1996,holevo,helstrom}. 
Each of these metrics induces a different geometry and, consequently, a different family of geodesics. 
In addition to these Fisher metric-based approaches, a very promising recent development is the study of the Bures-Wasserstein geometry, which provides a quantum analogue of optimal transport theory and is particularly suited to Gaussian states~\cite{villani2009,chen2017,amari2024}. 
This opens up exciting future research directions. Investigating the physical significance of geodesics corresponding to these different quantum geometries could provide new insights into the nature of quantum stochastic processes and thermalization. 
The choice of the ``correct" geometry may depend on the specific physical context, and exploring this connection is a promising area for future work.

\begin{acknowledgments}
T.K. thanks S.\ Amari for his insightful comments. 
T.K. acknowledges the financial support by CNPq (No.\ 304504/2024-6). A.vdV. is grateful for support from the Fueck-Stiftung.
A part of this work has been done under the project INCT-Nuclear Physics and Applications (No.\ 464898/2014-5).
\end{acknowledgments}

\end{document}